\documentclass[fleqn,usenatbib,useAMS]{mnras}

\usepackage{graphicx}	
\usepackage{amsmath}	
\usepackage{amssymb}	
\usepackage{multicol}        
\usepackage{bm}		
\usepackage{pdflscape}	
\usepackage{subfigure}
\usepackage{tablefootnote}

\usepackage[T1]{fontenc}
\usepackage{ae,aecompl}

\title[AstroSat Detection of a QPO at $\sim$ 42 Hz in Cyg X-2]{\textbf{AstroSat Detection of a Quasi-periodic Oscillation at $\sim42$} Hz in Cygnus X-2}
\author[Vanzarmawii Chhangte et al. ]{Vanzarmawii Chhangte$^{1}$\thanks{E-mail: zarichhangte0739@gmail.com}, Jayashree Roy$^{2}$, Ranjeev Misra$^{2}$, Lalthakimi Zadeng$^{1}$
\\
$^{1}$Mizoram University, Tanhril, Aizawl, Mizoram 796004, India\\
$^{2}$Inter-University Center for Astronomy and Astrophysics, Post Bag 4, Pune, Maharashtra 411007, India}

\begin{document}
\label{firstpage}
\pagerange{\pageref{firstpage}--\pageref{lastpage}}
\maketitle

\begin{abstract}
We report the results of AstroSat observations of Cygnus X-2 during February 2016. The source's power density spectrum generated using LAXPC data revealed the presence of a prominent Quasi-periodic Oscillation (QPO) at $\sim42$ Hz with broadband continuum noise at lower frequencies at $\sim10$ Hz. The large effective area of LAXPC at $\gtrsim$30 keV allowed for an unprecedented study of the energy dependence of the QPO and the broad noise continuum. The fractional r.m.s increases with energy, and its shape is similar for both the QPO and the continuum noise, suggesting a common radiative origin. However, while the QPO exhibits hard time lags, with the high energy photons lagging the low ones by a few milliseconds, the continuum noise shows the opposite behavior. The photon spectrum from SXT and LAXPC in $0.7-30$ keV band comprises the soft component from a disc and a hard Comptonized component from a hot corona. While the energy dependence of the r.m.s shows that the QPO and the continuum noise variability are dominated by the Comptonized component, the change in sign of the time-lag suggests that the dynamic origin of the QPO may be in the disk while the noise continuum may originate from the corona.\\

\end{abstract}





\section{Introduction}

X-ray binaries are a class of binary stars, so-called because they emit X-rays. The binary star comprises of a companion star and an accretor. These X-rays are produced by matter falling from the donor star to the accretor. The donors are usually normal stars, and accretors are collapsed stars that are compact, e.g., neutron stars (NS), white dwarf (WD), or black holes (BH). The X-ray binaries are classified into two classes according to the masses of the companion star- high-mass and low-mass X-ray binary.\\

The X-ray binary system comprising a NS as a compact object and a low-mass star as its companion is classified as a neutron star low-mass X-ray binary (NS LMXB). These can be sub-divided into Z-type and Atoll-type sources based on X-ray spectral and fast timing behaviour \citep{Hasinger1989}. The Z sources trace out a Z-shape in the X-ray color-color diagram (CCD). The branches of the Z-type sources are horizontal, normal, and flaring, from top to bottom. The Z sources are classified into two groups \citep{Kuulkers1997,Hasinger1989}, the Cyg-like sources- Cyg X-2, GX 5-1, GX 340+0 where the HB, NB, and FB are seen but with weak flaring and  the Sco-like sources- Sco X-1, GX 349+2, GX 17+2 where flaring is strong and frequent but with a short or weak HB.\\

A characteristic feature of these sources is their rapid nearly sinusoidal variation which is revealed by  peaks in their power spectra known as Quasi-periodic oscillations (QPO). In the horizontal branch (HB), the frequencies of the QPO vary between $\sim 15$Hz to $\sim 55$Hz \citep{AlparShaham1985,Lamb1989,Ghosh1992}, while in the normal branch (NB), the frequencies range between 5-7Hz and  in the flaring branch (FB), the frequencies are observed to increase from $\sim 6$Hz up to $\sim 20$Hz along the branch \citep{Hasinger1990}. The sources also exhibit high frequency kHz QPOs, which sometimes occur in pairs \citep{Wij1998,Kuznetsov2002}. Despite several endeavours to characterise the phenomena, there is at present no consensus on the origin of these different kinds of QPOs.\\

\citealt{Byram1966} first discovered  Cygnus X-2 (henceforth  Cyg X-2) using the sounding rocket experiment. It was first observed by EXOSAT for a continuous duration of 14h starting on July 23rd, 1984, using Gas scintillation proportional counter \citep{Peacock1981} and one-half of medium energy detectors \citep{Turner1981}. The X-ray binary source- Cyg X-2, is a bright, persistent LMXB. Cyg X-2 is classified as a Z-type source because of its behavior and pattern when studied on an X-ray color-color diagram (CCD) and hardness-intensity diagram (HID) \citep{Hasinger1989,Hasinger1990,Vander2000}. After the observation of thermonuclear X-ray bursts in Cyg X-2, its compact companion was identified as a NS with a low magnetic field \citep{Kahn1984,Smale1998}, its mass is measured as  $M_x>1.78\pm0.23M_\odot$ \citep{Orosz1999}. The binary system has a late-type companion, V1341 Cyg with an orbital period of $\sim 9.8$ days  \citep{Cowley1979,Casares1998} and a mass ranging between 0.4 and 0.7$M_\odot$ whose spectral type seem to vary from A5 to F2 \citep{Cowley1979}.\\

Quasi-periodic oscillations as well as broad continuum noise features have been extensively observed in Cyg X-2, by timing studies carried out using the proportional counter array (PCA) onboard the RXTE satellite \citep{Wij1998}.  Simultaneous detection of twin peaks at 500 and 860Hz and the highest single kHz QPO at 1007Hz were reported by \citep{Wij1998}. QPO at 18-50Hz in the HB and 5.6Hz QPO in the NB were also seen in Cyg X-2 \citep{Elsner1986,Hasinger1986,Wij1997,Kuulkers1999}. A 5Hz QPO was detected in Cyg X-2 by \citealt{Norris1986} during an 8hr observation of the source. The observation carried out by \citealt{Kuulkers1997} showed a similar QPO near $\sim 40$Hz in the  source.\\

The LAXPC \citep{Yadav2016b,Agrawal2017} instrument onboard the space observatory AstroSat \citep{Agrawal2006,Singh2014} has the advantage over RXTE by having a larger effective area at energies $\gtrsim 30$ keV. It has detected high frequency kHz QPO  LMXB \citep{Chauhan2017} at high energies and has shown its ability to provide unprecedented energy dependent timing features \citep{Yadav2016a,Misra2017}. The SXT instrument \citep{Singh2016,Singh2017} onboard provides simultaneous spectral coverage which has allowed for extensive spectro-timing studies of X-ray binaries \citep{Leahy2019,Roy2020}. Here, we report the timing and spectral analysis results of an AstroSat observation of Cygnus X-2 during which a $\sim 42$Hz QPO was observed.\\

Section 2 discussed the data reduction methodology of LAXPC and SXT, sections 2.1 and 2.2 discussed the spectral and timing analysis. Under section 3 we have the discussion.

\section{Observations and Data Reduction}
While a detailed study of the spectral and temporal evolution of Cyg X-2 using all the available AstroSat data will be presented elsewhere, here we focus on a 
a single observation (Obs ID 9000000348) undertaken from February 28th 2016 to March 1st 2016 with a start time of 52219.2985s and an exposure time of 64300s for LAXPC and 14290s for SXT.  Spectra were fitted using XSPEC package, version  12.10.1 \citep{Arnaud1996}.\\

The Large Area X-ray Proportional Counter (LAXPC) instrument abroad the AstroSat mission covers 3.0-80.0 keV broad spectral band for studying high time resolution and low spectral resolution of X-ray sources \citep{Agrawal2006,Antia2021}. The three LAXPC detectors have a total effective area of about $4500$cm$^2$ at 5keV, $6000$cm$^2$ at 10keV and about $5600$cm$^2$ at $\sim 40$keV \citep{Roy2019}. Details on the characteristics and calibrations of the LAXPC instruments can be found in \citep{Yadav2016b,Agrawal2017,Roy2016,Antia2017}. Data reduction was performed using LAXPCSOFT software (http://astrosat-ssc.iucaa.in/laxpcData, version as of August 04 2020) to reduce level-1 data to level-2, and standard routines in LAXPCSOFT were used to generate light curves, energy spectrum and background files. Since the observation was taken on 2016, we have included data from all the three LAXPC detectors. Later, one of the units, LAXPC 30. developed a gas leak and stopped operating on March 8 2018.\\
 
The soft X-ray imaging telescope (SXT) abroad AstroSat has been designed to provide soft X-ray images and spectra in the energy range 0.3-8.0keV by using conical foil mirrors for X-ray reflection with X-ray Charge Coupled Device (CCD) as the focal plane detector and details of the instrument are given in  \citet{Navalgund2017,Singh2014}. The effective area of SXT is $\sim$ 90cm$^2$ at 1.5keV. The SXT level 1 data were processed using SXTPIPELINE version AS1SXTLevel2-14b released on January 3rd 2019 to generate level-2 data for each orbit. The pipeline provides a Julia code to merge the level-2 SXT data of individual orbits. To avoid pile-up effect of SXT we select an annular region of 1 arcmin inner radius and 8 arcmin outer radius centered on the source to extract the image, light curves and spectra of the source. We have used SkyBkg\_comb\_EL3p5\_Cl\_Rd16p0\_v01.pha and sxt\_pc\_mat\_g0to12.rmf for background and rmf file respectively. Also the SXT arf generation tool sxtARFModule was used to generate vignetting corrected arf ARFTESTS1\_Rad4p0\_vigCorr.arf provided by SXT instrument team. The XSELECT utility from the HEASOFT package (v 6.26.1) was used to generate the spectrum. 
We have studied here a single observation (Obs ID 9000000348) of Cygnus X-2 observed during February-March 2016 by AstroSat. We have used data taken by both LAXPC and SXT. Analysis of both LAXPC and SXT were done using the LAXPCSoft software and XSELECT respectively. To create the merged event files from the clean event files, we have used the SXT event merger tool. Spectra were fitted using XSPEC 12.10.1.\\

\subsection{Spectral Analysis}

\begin{figure}
 \centering
    \includegraphics[scale=0.43,angle=270]{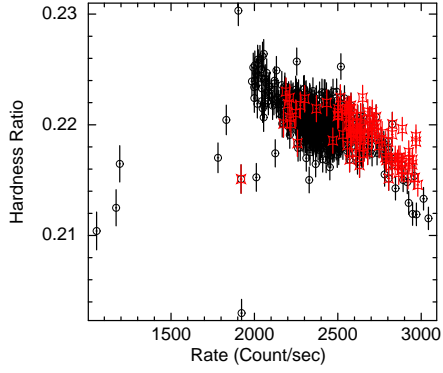}
    \caption{X-ray hardness intensity diagram of Cyg X-2. The horizontal branch of the Z pattern is seen. The black circle symbol represents the duration where only LAXPC data is available. Red square symbols indicate the duration when LAXPC and SXT data overlap.}
\label{f1}
\end{figure}

We have first examined the hardness intensity diagram (HID) constructed using LAXPC lightcurves binned in 100 sec and in the energy ranges $4$-$8$ keV (soft) and $8$-$10$keV (hard) as shown in Figure \ref{f1}. The Hardness defined as the ratio of the hard to soft bands does not show significant variation with the total count rate, which suggests that the source is in the Horizontal Branch. Similar variation is seen when different energy ranges are used.

A simultaneous broadband spectral analysis of SXT and LAXPC 20 was performed. LAXPC 20 was chosen since the LAXPC 20 data has less background compared to the others and since the spectrum is systematic dominated, including the other counters does not provide any significant improvement in the statistics. A gain correction was applied to the SXT spectrum using the Xspec command gain fit where the slope was fixed at unity and the best fit offset obtained was around $\sim 1.3$eV. A constant factor was included in the model to allow for cross calibration variation between SXT and LAXPC spectra. For LAXPC the constant faction was fixed at unity while for SXT this factor was allowed to vary. An overall systematic error of 3\% was included in the spectral fitting. For SXT the energy range considered for the fitting was $0.7$-$7.0$keV, while for LAXPC it was $4$-$30$keV, since beyond 30keV, the LAXPC spectrum was dominated by the background. 

\begin{figure}
 \centering
    \includegraphics[scale=0.41, angle=270]{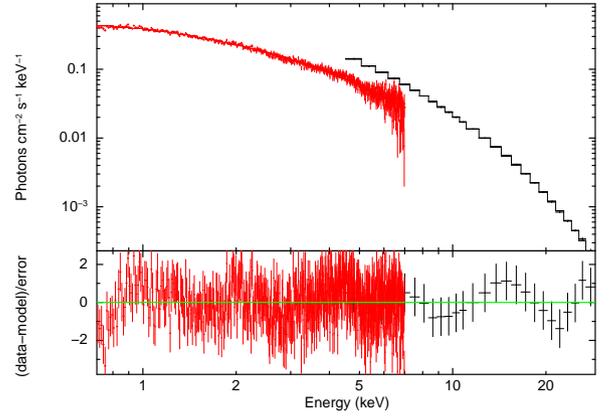}
    \caption{Spectra of LAXPC and SXT fitted with a model consisting of Comptonizing corona and disk emission with interstellar absorption with best fit parameters listed in Table \ref{tab:par2}. The black and red data points represent the LAXPC20 and SXT data respectively.}
    \label{f2}
\end{figure}

We fitted the combined spectra with a model consisting of interstellar absorption, a disc emission and a thermal Comptonized component represented by the XSPEC models {\it Tbabs} \citep{Wilms2000}, {\it diskbb} \citep{Mitsuda1984} and {\it thcomp} \citep{Zd2020}. Note that since {\it thcomp} is an convolution model, the model format used in XSPEC was {\it Tbabs*thcomp*diskbb} and we used the energies command, such that the model spectrum was computed for a wider energy range of 0.1-100 keV. The fitted spectra is shown in Fig. \ref{f2} and the best fit parameters with errors in the $90$\% confidence level are listed in Table \ref{tab:par2}.

While a detailed spectral evolution of the source with different models and using all the AstroSat observations will be presented elsewhere, here we point out that during this observation, the spectrum of Cyg X-2 was typical with a soft component represented by a disk with inner radius temperature $\sim 1.5$keV and thermal Comptonization from a corona with temperature $\sim 3.5$keV having an optical depth of $\sim 2.3$.

\begin{table}
	\centering
	\caption{X-ray Spectral Parameters of ObsID:9000000348 of Cyg X-2}
	\label{tab:par2}
	\begin{tabular}{lll}
	\hline
	Model Components & Parameters & Values\\
	\hline
	Tbabs & nH ($\times10{^{22}cm^{-2}}$) & $0.06_{-0.02}^{+0.01}$\\
	thcomp & $\Gamma$ & $2.3_{-0.5}^{+0.3}$\\
	& $kT_e$(keV) & $3.5_{-0.4}^{+0.4}$\\
	& $cov_{frac}$ & $0.7_{-0.4}^{+0.3}$\\
	diskbb & $kT_{in}$(keV) & $1.5_{-0.1}^{+0.1}$\\
	& $N_{disk}$ & $70.2_{-14.2}^{+22.8}$\\
	\hline
	$\chi^2$/dof=861.41/630\\
	\hline
	\end{tabular}
	
	\footnotesize{{\it{Note}}-(1) hydrogen column density in units of $\times10{^{22}cm^{-2}}$, (2) Thomson optical depth,  (3) corona electron temperature in keV, (4) covering fraction, (5) temperature at inner disk radius in keV and  (6) diskbb normalisation.}
       
	\end{table}

\subsection{Timing Analysis}

\begin{figure}
 \centering
    \includegraphics[scale=0.35,angle=270]{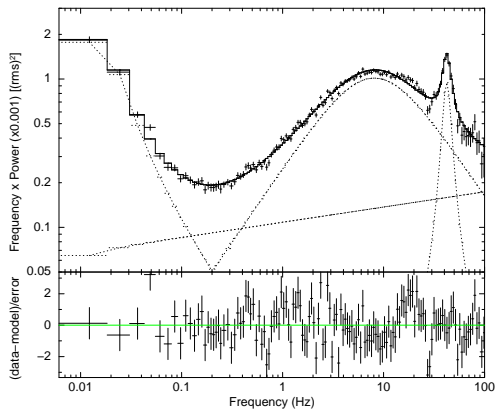}
    \caption{Power density spectrum of LAXPC from a single observation of Cyg X-2 in the 4.0-30.0keV fitted with three Lorentzians and a power law.}
    \label{f3}
\end{figure}

\begin{figure}
 \centering    
    \includegraphics[scale=0.4,angle=270]{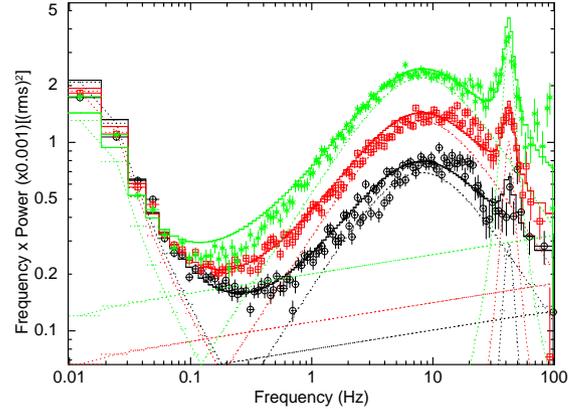}
   
    \caption{Power density spectra of LAXPC from a single observation of Cyg X-2 in the 4.0-6.0keV (black circle), 6.0-10.0keV (red box) and 10.0-30.0keV (green star) all fitted with three Lorentzians and a single powerlaw.}
    \label{f4}
\end{figure}

We generated the power density spectrum (PDS) from  the LAXPC data, using the
using the command `laxpc\_find\_freqlag' of the software LAXPCsoft Format A\footnote{http://astrosat-ssc.iucaa.in/laxpcData}. The software created the full lightcurve in the 4.0-30keV in 5 millisecond time-bin and divided it into 112 segments of 16384 time-bins. Power density spectrum for each segment was averaged and then rebbined in frequency. The resultant PDS in the frequency range 0.01-100Hz is shown in Fig. \ref{f3} which reveals a prominent peak at $\sim 42$Hz and broad band continuum noise. The high quality data shows complex broad band features, which  required three Lorentzian functions (one for the $\sim 42$Hz and two broad ones) and a power-law to fit the PDS even with a systematic error of 3$\%$. The best fit parameters are listed in Table \ref{tab:par1}. The residuals suggest that perhaps the empirical broad Lorentzian functions are not adequate to describe the continuum noise or that the shapes of the features evolve in time and hence the time averaged PDS is not being well represented. Nevertheless, a clear strong QPO at $\sim 42$ Hz and broad continuum is clearly evident in the data.  The  PDS were extracted in the same manner for  4.0-6.0, 6.0-10.0, 10.0-30.0keV bands and shown in Figure \ref{f4}. It is evident from the figure that the variability increases with energy.

\begin{figure}{ }
   \centering
\includegraphics[scale=0.6, angle=0]{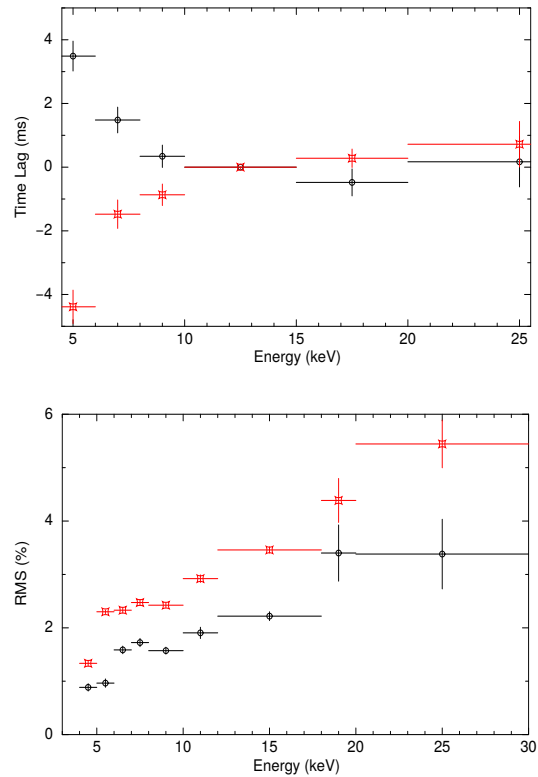}
 \caption{Time lag (top) and fractional r.m.s (bottom) and as a function of photon energy. For time-lag vs energy the black denotes the $\sim$10Hz while red denotes the $\sim$42Hz. The black denotes the $\sim$42 z while red denotes the $\sim$10Hz QPOs for the r.m.s vs energy.}
	\label{f5}

\end{figure}

The LAXPC software task `laxpc\_find\_freqlag' allows for the computation of the PDS for different energy bands, which is then used to estimate the Poisson noise subtracted and background flux corrected, fractional r.m.s, $frms$ for a frequency range $\Delta f$, as a function of energy. The code also generates using the cross-correlation function the energy dependent time-lag with respect to a reference energy band for a frequency range $\Delta f$. The top panel of Figure \ref{f5} shows the corresponding time-lags where the reference energy band is taken to be  $10$-$15$keV one. The bottom panel shows the energy dependent $frms$ for a $\Delta f\sim 3.9$Hz centered at $\sim 43$Hz, thus representing the QPO.  Also plotted in the figure are the $frms$ and time-lag corresponding to a $\Delta f \sim 3.9 $ centered at $\sim 11.7$Hz and hence representing the broad band continuum noise.

The fractional r.m.s increases with energy and has roughly the same shape for both the QPO and the continuum noise, indicating that most of the variability is in the higher energy thermal Comptonized spectral component. However, the energy dependent time-lag is qualitatively different for the QPO and the noise component. Negative time-lag at $\sim 4$keV with respect to $\sim 12$keV at the QPO frequency means that the high energy photons are delayed by about $\sim 4.5$milli-seconds compared to the low energy ones. On the other hand, for the continuum noise the opposite is seen, such that the low energy photons are delayed by $\sim 3.5$milli-seconds compared to the high energy ones. It should be noted that such a wide band energy dependence of the frms and time-lag has been possible due to the large effective area of the LAXPC instrument and  the strength of the QPO during this observation.

\begin{table}
\caption{Power Density Spectral Parameters for a single observation of Cyg X-2}
\label{tab:par1}
	\centering
	\begin{tabular}{lll}
	\hline
	Model Components & Parameters & Values\\
	\hline
		Lorentz & Centroid (Hz) 	& $42.0_{-0.4}^{+0.4}$\\               
		& Width (Hz) 		& $8.7_{-1.0}^{+1.1}$\\
		& rms (\%)			& $30.3_{-0.7}^{-0.8}$\\
		& Centroid (Hz) 		& $0_{0}^{+0.1}$\\
		& Width (Hz) 		& $16.1_{-0.3}^{+0.4}$\\
		& rms (\%)		& $11.4_{-0.2}^{+0.2}$\\
		& Centroid (milli-Hz) 	& $110_{-1}^{+1}$\\
		& Width (Hz) 		& $0.013_{-0.01}^{+0.01}$\\
		& rms (\%)		& $0.08_{-0.5}^{+0.5}$\\
	Powerlaw& PhoIndex 		& $0.9_{-0.04}^{+0.04}$\\
		& norm ($\times 10^{-4})$ & $1.1_{-0.08}^{+0.08}$\\
	\hline
\end{tabular}

\footnotesize{{\it{Note}}- PDS is fitted with three Lorentzian and a power law. All parameters are freed. $\chi^2$=184.68 dof=123}
\end{table}

\section{Discussion}

In thermal Comptonization, photons at different energies have scattered a different number of times inside the medium, which due to scattering time-scale can produce energy dependent time-lags. Indeed, this is the model invoked to explain the energy dependent r.m.s and time-lags observed for the kHz QPO in neutron star systems \citep{Lee2008,Kumar2014,Karpouzas2020}. The model predicts both hard and soft lags (the latter being observed in the lower kHz QPO), depending on the fraction of Comptonized photons that impinge back into the seed photon sources. The magntiude of the time lag is of the order of the light travel time through the corona and is measured to be $\sim 50$microseconds for the kHz QPO. The time-lags found in this work for the $\sim 42$Hz QPO are in the order of $\sim 5$milli-seconds and hence it is unlikely to be due to compton scattering effects, unless the size of the corona for this observation is $\sim 100$ times larger than when a khZ QPO is observed.\\

Time-lags significantly longer than those due to light travel time effects, can be explained in a framework where both the seed photon source and the Comptonizing medium coherently vary with a time-lag between them. As a specific example, a model where the heating rate of the corona and the temperature of the seed photon vary coherently with a time delay between them, can predict and fit the observed energy dependent r.m.s and time-lag for broad band noise and QPO of black hole systems \citep{Maqbool2019,Garg2020}. An obvious and generic feature of such models is that one gets hard lags when the coronal variation is delayed compared to that of soft photon source and the opposite soft lags occur when the the soft seed photon source variation occurs after the coronal one.\\

In this work, we find that while the broad band noise at $\sim 10$Hz exhibits soft lags, the $\sim 42$Hz QPO shows hard lags. The simplest explaination for this is to evoke causality and state that the broad band noise is generated in the corona and propogates to the soft photon source, while the QPO is generated in the soft photon source and then propogates to the corona.\\

If the soft photon source is taken to be an accretion disc surrounding the Comptonizing medium, it is interesting to note that the while the broad band noise is produced in the tenous corona, the narrow coherent QPO is produced in the disc, making it tempting to identify the QPO frequency with the inner disc radius. From the best fit value of the disk normalization (Table \ref{tab:par2}) $n \sim 70$, and using a distance, $D \sim 13.5$kpc \citep{Jonker2004}, inclination angle, $i \sim 60^\circ$ and colour factor, $f \sim 1.7$ \citep{Shimura1995}, the inner radius of the disc can be estimated to be $R_{in} \sim f^2 (D/10 kpc) \sqrt{70/cos(i)} \sim 45$kms. For a neutron star mass of $1.4M_\odot$, the Keplerian frequency at that radius is $\sim 240$Hz. The radius at which the Keplerian frequency would be equal to the QPO frequency of $\sim 42$Hz is $\sim 140$kms. Given the uncertainty in the spectral fitting, the spectral model used and the colour factor, it maybe possible that the QPO is associated with the Kelperian frequency of the inner disc radius. Alternatively it could be associated with one of the other lower frequencies characteristic of that radius. Nevertheless, the results of this work indicate that the QPO originates in the soft photon source and thus maybe associated with some characteristic time-scale of the inner disc radius.\\

\section{Data availability}
This research has made use of archival data from the \textit{AstroSat} mission.

\bibliographystyle{mnras}
\bibliography{ms}

\begin{thebibliography}{}
\makeatletter
\relax
\def\mn@urlcharsother{\let\do\@makeother \do\$\do\&\do\#\do\^\do\_\do\%\do\~}
\def\mn@doi{\begingroup\mn@urlcharsother \@ifnextchar [ {\mn@doi@}
  {\mn@doi@[]}}
\def\mn@doi@[#1]#2{\def\@tempa{#1}\ifx\@tempa\@empty \href
  {http://dx.doi.org/#2} {doi:#2}\else \href {http://dx.doi.org/#2} {#1}\fi
  \endgroup}
\def\mn@eprint#1#2{\mn@eprint@#1:#2::\@nil}
\def\mn@eprint@arXiv#1{\href {http://arxiv.org/abs/#1} {{\tt arXiv:#1}}}
\def\mn@eprint@dblp#1{\href {http://dblp.uni-trier.de/rec/bibtex/#1.xml}
  {dblp:#1}}
\def\mn@eprint@#1:#2:#3:#4\@nil{\def\@tempa {#1}\def\@tempb {#2}\def\@tempc
  {#3}\ifx \@tempc \@empty \let \@tempc \@tempb \let \@tempb \@tempa \fi \ifx
  \@tempb \@empty \def\@tempb {arXiv}\fi \@ifundefined
  {mn@eprint@\@tempb}{\@tempb:\@tempc}{\expandafter \expandafter \csname
  mn@eprint@\@tempb\endcsname \expandafter{\@tempc}}}

\bibitem[\protect\citeauthoryear{Agrawal}{Agrawal}{2006}]{Agrawal2006}
Agrawal P.,  2006, Adv.Space Res., 38, 2989

\bibitem[\protect\citeauthoryear{Agrawal et~al.}{Agrawal
  et~al.}{2017}]{Agrawal2017}
Agrawal P.,  et~al., 2017, J. Astrophys. Astron., 38, 30

\bibitem[\protect\citeauthoryear{Alpar \& Shaham}{Alpar \&
  Shaham}{1985}]{AlparShaham1985}
Alpar M.~A.,  Shaham J.,  1985, Nat, 316, 239

\bibitem[\protect\citeauthoryear{Antia et~al.}{Antia et~al.}{2017}]{Antia2017}
Antia H.~M.,  et~al., 2017, ApJS, 231, 1

\bibitem[\protect\citeauthoryear{Antia et~al.}{Antia et~al.}{2021}]{Antia2021}
Antia H.~M.,  et~al., 2021, J. Astrophys. Astron., 42, 32

\bibitem[\protect\citeauthoryear{Arnaud}{Arnaud}{1996}]{Arnaud1996}
Arnaud H.,  1996, Astronomical Data Analysis Software and Systems V. Astron.
  Soc. Pac, 101, 17

\bibitem[\protect\citeauthoryear{Byram, Chubb  \& Friedman}{Byram
  et~al.}{1966}]{Byram1966}
Byram E.~T.,  Chubb T.~A.,   Friedman H.,  1966, Science, 152, 66

\bibitem[\protect\citeauthoryear{Casares, Charles  \& Kuulkers}{Casares
  et~al.}{1998}]{Casares1998}
Casares J.,  Charles P.,   Kuulkers E.,  1998, ApJ, 493, 1

\bibitem[\protect\citeauthoryear{Chauhan et~al.}{Chauhan
  et~al.}{2017}]{Chauhan2017}
Chauhan J.~V.,  et~al., 2017, ApJ, 841, 41

\bibitem[\protect\citeauthoryear{Cowley, Crampton  \& Hutchings}{Cowley
  et~al.}{1979}]{Cowley1979}
Cowley A.~P.,  Crampton D.,   Hutchings J.~B.,  1979, ApJ, 231, 539

\bibitem[\protect\citeauthoryear{Elsner et~al.}{Elsner
  et~al.}{1986}]{Elsner1986}
Elsner R.~F.,  et~al., 1986, ApJ, 308, 655

\bibitem[\protect\citeauthoryear{Garg, Misra  \& Sen}{Garg
  et~al.}{2020}]{Garg2020}
Garg A.,  Misra R.,   Sen S.,  2020, MNRAS, 499, 2757

\bibitem[\protect\citeauthoryear{Ghosh \& Lamb}{Ghosh \&
  Lamb}{1992}]{Ghosh1992}
Ghosh P.,  Lamb F.~K.,  1992, In: van den Heuvel E.P.J., Rappaport S.A. (eds)
  X-Ray Binaries and Recycled Pulsars. NATO ASI Series (Series C: Mathematical
  and Physical Sciences), 377

\bibitem[\protect\citeauthoryear{Hasinger}{Hasinger}{1990}]{Hasinger1990}
Hasinger G.,  1990, Rev. Mon. Astron., 3, 60

\bibitem[\protect\citeauthoryear{Hasinger \& van~der Klis}{Hasinger \& van~der
  Klis}{1989}]{Hasinger1989}
Hasinger G.,  van~der Klis M.,  1989, A\&A, 225, 79

\bibitem[\protect\citeauthoryear{Hasinger et~al.}{Hasinger
  et~al.}{1986}]{Hasinger1986}
Hasinger G.,  et~al., 1986, Nat, 319, 469

\bibitem[\protect\citeauthoryear{Jonker \& Nelemans}{Jonker \&
  Nelemans}{2004}]{Jonker2004}
Jonker P.~G.,  Nelemans G.,  2004, MNRAS, 354, 355

\bibitem[\protect\citeauthoryear{Kahn \& Grindlay}{Kahn \&
  Grindlay}{1984}]{Kahn1984}
Kahn S.~M.,  Grindlay J.~E.,  1984, ApJ, 281, 826

\bibitem[\protect\citeauthoryear{Karpouzas et~al.}{Karpouzas
  et~al.}{2020}]{Karpouzas2020}
Karpouzas K.,  et~al., 2020, MNRAS, 492, 1399

\bibitem[\protect\citeauthoryear{Kumar \& Misra}{Kumar \&
  Misra}{2014}]{Kumar2014}
Kumar N.,  Misra R.,  2014, MNRAS, 445, 2818

\bibitem[\protect\citeauthoryear{Kuulkers et~al.}{Kuulkers
  et~al.}{1997}]{Kuulkers1997}
Kuulkers E.,  et~al., 1997, MNRAS, 287, 495

\bibitem[\protect\citeauthoryear{Kuulkers, Wijnands  \& Van Der~Klis}{Kuulkers
  et~al.}{1999}]{Kuulkers1999}
Kuulkers E.,  Wijnands R.,   Van Der~Klis M.,  1999, MNRAS, 308, 485

\bibitem[\protect\citeauthoryear{Kuznetsov}{Kuznetsov}{2002}]{Kuznetsov2002}
Kuznetsov S.~I.,  2002, Astron. Lett, 28, 73

\bibitem[\protect\citeauthoryear{Lamb}{Lamb}{1989}]{Lamb1989}
Lamb F.~K.,  1989, In: Ogelman H., van den Heuvel E.P.J. (eds) Timing Neutron
  Stars. NATO ASI Series (Series C: Mathematical and Physical Sciences), 262

\bibitem[\protect\citeauthoryear{Leahy \& Chen}{Leahy \&
  Chen}{2019}]{Leahy2019}
Leahy D.~A.,  Chen Y.,  2019, ApJ, 871, 152

\bibitem[\protect\citeauthoryear{Lee, Misra  \& Taam}{Lee
  et~al.}{2008}]{Lee2008}
Lee H.,  Misra R.,   Taam R.,  2008, ApJ, 549, L229

\bibitem[\protect\citeauthoryear{Maqbool et~al.}{Maqbool
  et~al.}{2019}]{Maqbool2019}
Maqbool B.,  et~al., 2019, MNRAS, 486, 2964

\bibitem[\protect\citeauthoryear{Misra et~al.}{Misra et~al.}{2017}]{Misra2017}
Misra R.,  et~al., 2017, ApJ, 835, 195

\bibitem[\protect\citeauthoryear{Mitsuda et~al.}{Mitsuda
  et~al.}{1984}]{Mitsuda1984}
Mitsuda K.,  et~al., 1984, PASJ, 36, 741

\bibitem[\protect\citeauthoryear{Navalgund et~al.}{Navalgund
  et~al.}{2017}]{Navalgund2017}
Navalgund K.~H.,  et~al., 2017, J. Astrophys. Astron., 38, 34

\bibitem[\protect\citeauthoryear{Norris \& Wood}{Norris \&
  Wood}{1986}]{Norris1986}
Norris J.~P.,  Wood K.~S.,  1986, ApJ, 312, 732

\bibitem[\protect\citeauthoryear{Orosz \& Kuulkers}{Orosz \&
  Kuulkers}{1999}]{Orosz1999}
Orosz J.~A.,  Kuulkers E.,  1999, MNRAS, 305, 132

\bibitem[\protect\citeauthoryear{Peacock et~al.}{Peacock
  et~al.}{1981}]{Peacock1981}
Peacock A.,  et~al., 1981, In: Andresen R.D. (eds) X-Ray Astronomy. Springer,
  Dordrecht.

\bibitem[\protect\citeauthoryear{Roy et~al.}{Roy et~al.}{2016}]{Roy2016}
Roy J.,  et~al., 2016, ApJ, 42, 249

\bibitem[\protect\citeauthoryear{Roy et~al.}{Roy et~al.}{2019}]{Roy2019}
Roy J.,  et~al., 2019, ApJ, 872, 33

\bibitem[\protect\citeauthoryear{Roy et~al.}{Roy et~al.}{2020}]{Roy2020}
Roy J.,  et~al., 2020, Res. Astron. Astrophys, 20, 155

\bibitem[\protect\citeauthoryear{Shimura \& Takahara}{Shimura \&
  Takahara}{1995}]{Shimura1995}
Shimura T.,  Takahara F.,  1995, ApJ, 445, 780

\bibitem[\protect\citeauthoryear{Singh et~al.}{Singh et~al.}{2014}]{Singh2014}
Singh K.~P.,  et~al., 2014, Proc. SPIE, 9144, 517

\bibitem[\protect\citeauthoryear{Singh et~al.}{Singh et~al.}{2016}]{Singh2016}
Singh K.~P.,  et~al., 2016, SPIE, 9905, 99051E

\bibitem[\protect\citeauthoryear{Singh et~al.}{Singh et~al.}{2017}]{Singh2017}
Singh K.~P.,  et~al., 2017, JApA, 38, 29

\bibitem[\protect\citeauthoryear{Smale}{Smale}{1998}]{Smale1998}
Smale A.~P.,  1998, ApJ, 498, 141

\bibitem[\protect\citeauthoryear{Turner, Smith  \& Zimmermann}{Turner
  et~al.}{1981}]{Turner1981}
Turner M. J.~L.,  Smith A.,   Zimmermann H.~U.,  1981, In: Andresen R.D. (eds)
  X-Ray Astronomy. Springer, Dordrecht., pp 513--524

\bibitem[\protect\citeauthoryear{Van~der Klis}{Van~der Klis}{2000}]{Vander2000}
Van~der Klis M.,  2000, ARA\&A, 38, 717

\bibitem[\protect\citeauthoryear{Wijnands et~al.}{Wijnands
  et~al.}{1997}]{Wij1997}
Wijnands R. A.~D.,  et~al., 1997, A\&A, 323, 399

\bibitem[\protect\citeauthoryear{Wijnands et~al.}{Wijnands
  et~al.}{1998}]{Wij1998}
Wijnands R. A.~D.,  et~al., 1998, ApJ, 493, 87

\bibitem[\protect\citeauthoryear{Wilms, Allen  \& McCray}{Wilms
  et~al.}{2000}]{Wilms2000}
Wilms J.,  Allen A.,   McCray R.,  2000, ApJ, 542, 914

\bibitem[\protect\citeauthoryear{Yadav et~al.}{Yadav
  et~al.}{2016a}]{Yadav2016a}
Yadav J.~S.,  et~al., 2016a, ApJ, 833, 27

\bibitem[\protect\citeauthoryear{Yadav et~al.}{Yadav
  et~al.}{2016b}]{Yadav2016b}
Yadav J.~S.,  et~al., 2016b, Proc. SPIE, 9905, 374

\bibitem[\protect\citeauthoryear{Zdziarski et~al.}{Zdziarski
  et~al.}{2020}]{Zd2020}
Zdziarski A.~A.,  et~al., 2020, MNRAS, 492, 5234

\makeatother
\end{thebibliography}

\end{document}